\documentclass[12pt]{article}
\usepackage{amsmath,bbold,amsfonts,amssymb,graphics,xcolor}
\usepackage[T1]{fontenc}
\usepackage{graphicx}
\usepackage{tikz}

\makeatletter

\def\TToverline{[T\overline{T}]}

\def\uqsu2{\mathcal{U}_{\mathsf{q}}(\mathfrak{su}_2)}

\def\half{{1\over{2}}}

\newcommand{\beq}{\begin{equation}}
\newcommand{\eeq}{\end{equation}}
\newcommand{\bea}{\begin{eqnarray}}
\newcommand{\eea}{\end{eqnarray}}
\newcommand{\beaq}{\begin{eqnarray}}
\newcommand{\eeaq}{\end{eqnarray}}
\begin{document}
\begin{flushright}\footnotesize
\end{flushright}
\vspace{0.7cm}
\centerline{\Large \bf Massless $S$-matrix Bootstraps and RG Flows}
\vskip 1cm
\centerline{\large Changrim Ahn\footnote{ahn@ewha.ac.kr}  }
\vskip 1cm
\centerline{\it Department of Physics}
\centerline{\it Ewha Womans University}
\centerline{\it Seoul 120-750, Korea}
\centerline{\small PACS: 11.25.Hf, 11.55.Ds}
\vskip 0.8cm
\centerline{\bf Abstract}
We find complete solutions of $S$-matrix bootstrap equations for the scatterings of massless particles in (1+1) dimensions with $A_n$ symmetries. We show that only three types (minimal, diagonal, and saturated cases) of these $S$-matrices can generate RG flows that lead to UV-complete theories, and other RG flows predicted in \cite{AhnLeC} are inconsistent with the bootstrap equations. 
Using these $S$-matrices, we derived the TBA equations and corresponding $Y$-systems that generate the RG flows between the IR and UV CFTs.

\newpage

\section{Introduction}

Integrable quantum field theories provide quantitative methods for non-perturbative computations.
A traditional approach  is to construct the integrability from a ultraviolet (UV) conformal field theory (CFT) perturbed by a relevant operator and derive 
exact $S$-matrices between massive on-shell particles using symmetries and other axioms, as
described in \cite{sasha}.
The $S$-matrices play an essential role in both defining the theories and computing physical quantities.
Thermodynamic Bethe ansatz (TBA) equations \cite{alyosha_tba} derived from these exact $S$-matrices 
can provide the UV data of the perturbed CFT, such as the central charges and dimensions of the relevant operators.
Among these integrable QFTs, those with UV  and infrared (IR)  fixed points are of special interest.
These can show how states in both fixed points are connected by renormalization group (RG) flows.
However, finding and solving such QFTs are still quite challenging because the theory becomes non-perturbative 
in the IR scale.
Traditional approaches are based on perturbative CFT computations \cite{zamolRG}, conjectured TBAs \cite{alyosha_tba_flow, alyosha_tba_cosetflow, Ravanini,DDT}, and efforts to find massless $S$-matrices from which exact TBAs can be derived \cite{zamzam_flow, FSZ, AKRZ}.
Still, some TBA systems remain as conjectures since$S$-matrices are not discovered so that 
exact computations of other interesting quantities are unavailable.

A recent approach to address this issue was proposed in \cite{AhnLeC}.
The basic idea is to consider a perturbed CFT with additional deformations by higher energy-momentum tensors 
\beq
\mathcal{S}_{\alpha}=\mathcal{S}_{{\rm cft}_{\rm IR}}+\lambda\int d^2x\Phi_{\rm rel}+\sum_{s\ge 1}\alpha_s\int d^2x \TToverline_s.
\label{pertcftTT}
\eeq
In the massless limit $\lambda\to 0$, 
$S$-matrices, $S^{LL}$, of two left-moving ($L$) particles  and $S^{RR}$ of two right-moving ($R$) particles
remain the same as the already known massive $S$-matrices. 
In contrast, the scatterings between $R$ and $L$ particles, $S^{RL}$, receive non-trivial contributions from the massive CDD factors prescribed by a seminal paper \cite{SmiZam} as follows:
\beq
S^{RL}_{ab}(\theta)=S^{LR}_{ab}(-\theta)=\exp\left(i\sum_{s\ge 1}g^{ab}_s\,e^{s\theta}\right)
\label{SabTT}
\eeq
where indices $a,b$ denote particle species with masses $m_a$ and $\theta$ is the rapidity difference of the two particles.
The coefficients $g^{ab}_s$ in \eqref{SabTT} are related to $\alpha_s$ in Eq.\eqref{pertcftTT}.

If the $S^{RR}$-matrix is non-diagonal, the $S^{RL}$-matrix should satisfy the Yang-Baxter equation, which imposes strong constraints on it \cite{AhnBaj}.
For diagonal scatterings, which we will consider in this study, the expressions in Eq.\eqref{SabTT} are too broad.
We need to find additional conditions on the $S^{RL}$-matrices.

In \cite{AhnLeC}, these were imposed in two steps.
The first is the crossing-unitarity relations
\beq
S^{RL}_{ab}(\theta)S^{RL}_{ab}(i\pi+\theta)=1,
\label{neutralcrossunit}
\eeq
which are satisfied if $S^{RL}_{ab}(\theta)$ are products of two basic solutions of \eqref{neutralcrossunit} 
\beq
T(\theta)=\tanh\frac{1}{2}\left(\theta-\frac{i\pi}{2}\right),\qquad F_{\gamma}=\frac{\sinh\theta-i\sin\pi\gamma}{\sinh\theta+i\sin\pi\gamma}.
\label{neutralsols}
\eeq
A further restriction comes from the claim that the central charges of UV-complete theories should be finite and rational numbers.
This has been imposed on the plateau equations, whose exponents are determined by the numbers of $T$ and $F_{\gamma}$ functions in each $S^{RL}_{ab}$.
Solving these algebraically coupled equations allows one to calculate the central charges, which can identify UV-complete theories. 
Several new solutions with rational central charges have been found in this manner.

This approach provides an effective way to classify all possible UV-complete theories for a given IR CFT.
However, the number of possible UV theories is rapidly increasing as scattering theories become more complex.
For example, only three UV CFTs can be associated with the simplest $su(3)$ coset CFT by RG flows.
For a $su(4)$ coset CFT, the number of UV CFTs is 11. 
A natural question is whether all these solutions are correct UV-complete theories.
To answer this, we need to derive and analyse the full TBA systems from which 
we can identify the deforming relevant field of each UV-complete theory.

In this paper, we address these issues using massless bootstrap equations and crossing-unitarity relations. 
Although we consider only the $A_n$ coset CFTs here, our method can be applied to other simply laced Lie algebras.
\section{Massless Bootstrap equations for $A_n$ theories}\label{sec2}
Here, we will determine all possible $S$-matrices from massless $S$-matrix bootstrap equations and crossing-unitarity relations.

\subsection{Massive scattering theories}
We start with a coset CFT
\beq
[G]_k=\frac{G_1\times G_k}{G_{k+1}},
\label{cosetcft}
\eeq
where $G_k$ stands for the WZW CFT based on a simply laced Lie group $G$ and a level $k$ with central charge
$c_k=k{\rm dim}G/(k+h)$ ($h$ is the dual Coxeter number).
The central charges of this series are given by $c([G]_k)=c_{1}+c_{k}-c_{k+1}$.

It is known that this CFT deformed by a relevant coset field is integrable
\beq
\mathcal{S}_{\lambda}=\mathcal{S}_{[G]_k}+\lambda\int d^2x \Phi^{(0)}_{{\rm rel};k},\qquad
\Phi^{(0)}_{{\rm rel};k}=\left[\frac{(1;\circ)]\times (k;\circ)}{(k+1;{\rm Adj})}\right],
\label{deformedcft}
\eeq
where $\circ,\ {\rm Adj}$ denote the singlet and adjoint representations of the group $G$, respectively.
The dimension of the relevant field is
\beq
\Delta\left(\Phi^{(0)}_{{\rm rel};k}\right)=\frac{2(k+1)}{k+1+h}.
\label{dimrel}
\eeq

For negative $\lambda$, the particle spectrum is massive, and its exact $S$-matrices are given by the quantum group restrictions of the affine Toda field theory for the group $G$ \cite{ABL}.
For $\lambda>0$, this model is conjectured to generate massless flows from UV CFT $[G]_k$ to IR CFT $[G]_{k-1}$
along its least irrelevant operator \cite{Ravanini}
\beq
\Phi^{(0)}_{{\rm irrel};k-1}=\left[\frac{(1;\circ)]\times (k-1;{\rm Adj})}{(k;\circ)}\right]
\label{irrel}
\eeq
which has dimension $2(1+h/(k-1+h))$.
If $k=2$, the leading irrelevant operator guiding the flow from $[G]_2$ to $[G]_1$ should be 
the $\TToverline$ since the irrelevant operator \eqref{irrel} cannot exist.

For the simplicity of arguments, we will consider $G=A_n$ with arbitrary rank $n$.
These theories are described by particle spectrum $A_a,\ a=1,\cdots, n$ and 
two-particle scattering amplitudes $S_{ab}$, the $S$-matrix, defined by
\beq
A_a(\theta_1)A_b(\theta_2)=S_{ab}(\theta_1-\theta_2)A_b(\theta_2)A_a(\theta_1).
\label{defS}
\eeq
According to the $S$-matrix bootstrap program \cite{sasha}, 
a particle $A_c$ can be a bound state of two particles $A_a$ and $A_b$, namely,
\beq
\vert A_{\overline c}(\theta)\rangle=\vert A_a(\theta+i\overline{u}_{ac}^b)\cdot A_b(\theta-i\overline{u}_{bc}^a)\rangle.
\label{bound}
\eeq
Here, $\overline{c}$ represents the antiparticle of particle $c$. 
If the particles are neutral, $\overline{c}=c$.
When this bound state scatters with another particle $A_d$, the $S$-matrices should satisfy
\beq
S_{d\overline c}(\theta)=S_{da}(\theta-i\overline{u}_{ac}^b)\,S_{db}(\theta+i\overline{u}_{bc}^a),
\label{Sbootstrap}
\eeq
which we will refer to as $S$-matrix bootstrap equations.
Another condition is that the $S_{ab}(\theta)$ should have a simple pole of bound state $A_c$
with positive residue at $\theta=iu_{ab}^c$ with $0<u_{ab}^c<\pi$ ($\overline{u}_{ac}^b=\pi-u_{ac}^b$).
The mass of $A_c$ is given by $m_c=M\sin u_{ab}^c$.
In this way, the parameters $u_{ab}^c$ are determined in a self-consistent manner.

For $G=A_n$, each particle $A_c$ is not neutral but has anti-particle 
$A_{\overline{c}}$ where
\beq
\overline{c}=n+1-c.
\label{antipar}
\eeq
The parameters $u_{ab}^c$ are known for all ADE algebras \cite{BCDS}.
For $A_n$, they are given as follows:
\beq
u_{ab}^c=
\begin{cases}
\frac{a+b}{n+1}\pi& a+b+c=n+1\\
\frac{\overline{a}+\overline{b}}{n+1}\pi& a+b+c=2(n+1).
\end{cases}
\label{uabc}
\eeq
The solutions of \eqref{Sbootstrap} for massive deformed CFTs \eqref{deformedcft} with $G=A_n$ are given by 
\beq
S_{ab}(\theta)=(a+b)_{-}(a+b-2)_{-}^2\cdots (|a-b|-2)_{-}^2 (|a-b|)_{-},
\label{massiveAnS}
\eeq
where we use a short notation
\beq
(k)=\frac{\sinh\left[\half\left(\theta-\frac{k\pi}{h}\right)\right]}{\sinh\left[\half\left(\theta+\frac{k\pi}{h}\right)\right]},\qquad
(k)_{-}=(-k)=\frac{1}{(k)}.
\label{defbasicblock}
\eeq
From this definition, it is obvious that $(h)=-1,\ (k-2h)=(k)$ where $h=n+1$ is the dual Coxeter number of $A_n$.

These $S$-matrices also satisfy the parity and the charge conjugation symmetries,
\beq
S_{ab}(\theta)=S_{ba}(\theta)=S_{\overline{a}\overline{b}}(\theta), 
\label{PC}
\eeq
as well as the crossing symmetry and unitarity conditions
\beq
S_{a\overline{b}}(\theta)=S_{ab}(i\pi-\theta),\quad
S_{ab}(\theta)S_{ab}(-\theta)=1.
\label{crossNunit}
\eeq

This set of the massive $S$-matrices become the $S^{LL}$ and $S^{RR}$ which describe the 
IR CFT $[A_n]_1$ as we will show shortly.

\subsection{Massless Bootstrap equations and Crossing-Unitarity}

The massless limit $\lambda\to 0$ restores the conformal symmetry in Eq.\eqref{deformedcft}.
This limit can be imposed on the particle spectrum by  shifting the rapidity $\theta\to \theta\pm\theta_0$ 
where a double limit $\theta_0\to\infty,\ M\to 0$ produces the massless dispersion relation
$E_a=\pm P_a=\mu_a\kappa e^{\pm\theta}$  with $\kappa=Me^{\theta_0}$ finite.
Depending on the sign, two sets of particles arise:
the right-moving ($R$) particles $A^R_a(\theta)$ with an upper sign ($+$) and
the left-moving ($L$) particles $A^L_a(\theta)$
with a lower sign ($-$) for $a=1,\cdots,n$.

The bound state conditions \eqref{bound} are invariant under this shift,
\beq
\vert A^{R}_{\overline c}(\theta)\rangle=\vert A^{R}_a(\theta+i\overline{u}_{ac}^b)\, A^{R}_b(\theta-i\overline{u}_{bc}^a)\rangle,
\quad
\vert A^{L}_{\overline c}(\theta)\rangle=\vert A^{L}_a(\theta+i\overline{u}_{ac}^b)\, A^{L}_b(\theta-i\overline{u}_{bc}^a)\rangle.
\label{boundRL}
\eeq
Therefore, the $S$-matrix bootstrap equations for the $S_{ab}^{LL}$ and $S_{ab}^{RR}$ are the same as those for the massive case \eqref{massiveAnS}. 
The shift $\theta_0$ cancels in the difference of the rapidities.

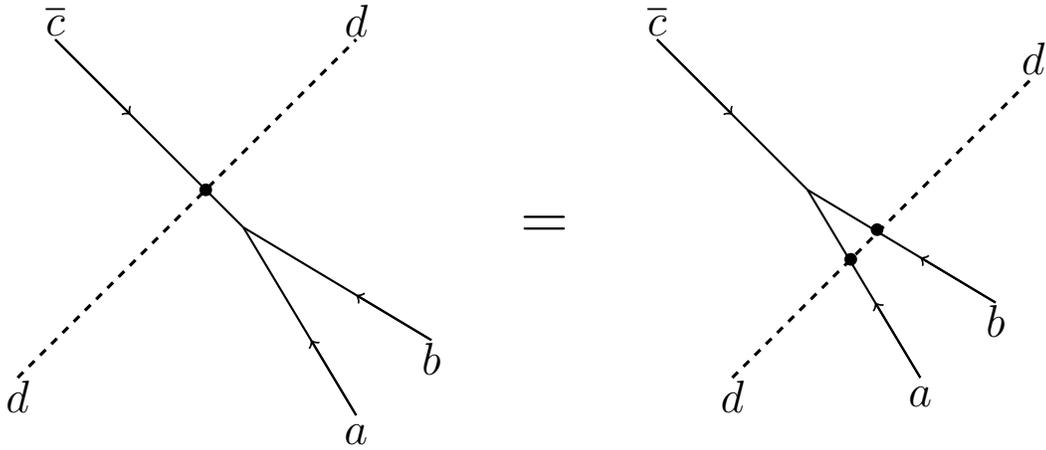
\begin{figure}[h]
\centering
\begin{tikzpicture}[scale=1, transform shape]
\draw[very thick,dashed] (-1*0.5,1*0.5) -- (8*0.5,10*0.5);
\draw[thick] (8*0.5,0) -- (5*0.5,5*0.5);
\draw[thick,->] (8*0.5,0) -- (8*0.5-1.2*0.5,0+2*0.5);
\draw[thick] (10*0.5,2*0.5) -- (10*0.5-5*0.5,2*0.5+3*0.5);
\draw[thick,->] (10*0.5,2*0.5) -- (10*0.5-2*0.5,2*0.5+1.2*0.5);
\draw[thick] (5*0.5,5*0.5) -- (0,10*0.5);
\draw[thick,->] (0,10*0.5) -- (0+2*0.5,10*0.5-2*0.5);
\filldraw[color=black, fill=black, thick] (4*0.5,6*0.5) circle (2pt);
\draw[very thick,dashed] (-2*0.5+20*0.5,0+1*0.5) -- (8*0.5+18*0.5,8*0.5+1*0.5);
\draw[thick] (8*0.5+15*0.5,0+1*0.5) -- (5*0.5+15*0.5,5*0.5+1*0.5);
\draw[thick,->] (8*0.5+15*0.5,0+1*0.5) -- (8*0.5+15*0.5-1.2*0.5,2*0.5+1*0.5);
\draw[thick] (10*0.5+15*0.5,2*0.5+1*0.5) -- (5*0.5+15*0.5,5*0.5+1*0.5);
\draw[thick,->] (10*0.5+15*0.5,2*0.5+1*0.5) -- (10*0.5+15*0.5-2*0.5,2*0.5+1*0.5+1.2*0.5);
\draw[thick] (5*0.5+15*0.5,5*0.5+1*0.5) -- (5*0.5+15*0.5-4*0.5,6*0.5+4*0.5);
\draw[thick,->] (16*0.5,10*0.5) -- (18*0.5,8*0.5);
\filldraw[color=black, fill=black, thick] (4*0.5+17.5*0.5+0.35*0.5,3.5*0.51+0.35*0.5+1*0.51) circle (2pt);
\filldraw[color=black, fill=black, thick] (4*0.5+17.5*0.5-0.35*0.5,3.5*0.5-0.35*0.5+1*0.5) circle (2pt);
\node [label] at (8*0.5,-0.5*0.5) {\Large $a$};
\node [label] at (10*0.5,1.5*0.5) {\Large $b$};
\node [label] at (-1*0.5,0.5*0.5) {\Large $d$};
\node [label] at (8*0.5,10.5*0.5) {\Large $d$};
\node [label] at (0,10.5*0.5) {\Large $\overline{c}$};
\node [label] at (23*0.5,0.5*0.5) {\Large $a$};
\node [label] at (25*0.5,2.5*0.5) {\Large $b$};
\node [label] at (18*0.5,0.5*0.5) {\Large $d$};
\node [label] at (26*0.5,9.5*0.5) {\Large $d$};
\node [label] at (16*0.5,10.5*0.5) {\Large $\overline{c}$};
\node [label] at (13*0.5,5*0.5) {\huge $=$};
\end{tikzpicture}
\caption{Graphical representation of the $S^{RL}$ bootstrap equations: Dotted (solid) lines denote $R(L)$-particles and 
upward (downward) arrows denote (anti)particles. }
\label{fig0}
\end{figure}

Now we define a scattering $S^{LR}$ by
\beq
A^{R}_{d}(\theta_1)A^{L}_{\overline c}(\theta_2)=S^{RL}_{d\overline c}(\theta_1-\theta_2)
A^{L}_{\overline c}(\theta_2)A^{R}_{d}(\theta_1),
\label{SRLdef}
\eeq
and use Eq.\eqref{boundRL} to get the massless $S$-matrix bootstrap equations
(Fig.\ref{fig0})
\beq
S^{RL}_{d\overline c}(\theta)=S^{RL}_{da}(\theta-i\overline{u}_{ac}^b)\,S^{RL}_{db}(\theta+i\overline{u}_{bc}^a),
\label{SbootstrapLR}
\eeq
and similarly for $S^{LR}$. 

One important observation is that $S_{ab}^{RL}$ should be {\it without any physical poles} 
because the $R$ and $L$ particles do not form any bound states.
This is different from the massive solutions like \eqref{massiveAnS} given by products of $(-k)$ with  $k$ positive integers, and hence has a pole at $\theta=ik\pi/h$.
Owing to this difference, more solutions can be found because we need not consider the poles of the $S_{ab}^{RL}$.

In addition to these bootstrap equations, we need to generalize the crossing-unitarity relation in \eqref{neutralcrossunit}, which is valid only for neutral particles. 
For $G=A_n$, two particles $A^{R/L}_a$ and $A^{R/L}_{\overline{a}}$ form charge conjugation pairs.
The general crossing-unitarity relation is
\beq
S_{ab}^{RL}(\theta)S_{\overline{a}b}^{RL}(\theta+i\pi)=1.
\label{crossunit}
\eeq


\begin{figure}[h]
\centering
\begin{tikzpicture}[scale=.7, transform shape]
\draw[thick] (1,0) -- (3,0);
\draw[thick] (3,0) -- (5,0);
\draw[thick] (7,0) -- (9,0);
\draw[thick] (9,0) -- (11,0);

\draw[thick,dashed] (5,0) -- (7,0);

\filldraw[color=black, fill=white, thick] (1,0) circle (4pt);
\filldraw[color=black, fill=white, thick] (3,0) circle (4pt);
\filldraw[color=black, fill=white, thick] (5,0) circle (4pt);
\filldraw[color=black, fill=white, thick] (7,0) circle (4pt);
\filldraw[color=black, fill=white, thick] (9,0) circle (4pt);
\filldraw[color=black, fill=white, thick] (11,0) circle (4pt);
\node [label] at (1,0.5) {$1$};
\node [label] at (3,0.5) {$2$};
\node [label] at (5,0.5) {$3$};
\node [label] at (11,0.5) {$n$};
\node [label] at (9,0.5) {$n-1$};
\node [label] at (7,0.5) {$n-2$};
\end{tikzpicture}
\caption{Dynkin diagram of $A_n$ algebra.}
\label{fig1}
\end{figure}
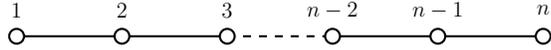

We will solve Eqs.\eqref{SbootstrapLR} and \eqref{crossunit} systematically to find all possible 
$S_{ab}^{RL}$ and UV-complete theories associated with them. 
From $u^{c}_{ab}$ in \eqref{uabc}, it is obvious that \eqref{SbootstrapLR} is satisfied if $a+b=c$ and $c\le n/2$.
If we choose an ansatz for $S^{RL}_{11}(\theta)$, all $S^{RL}_{1b}$ with $b=2,\cdots,n$ can be determined with $d=1$.
And from the $S^{RL}_{12}$, $S^{RL}_{22}$ can be obtained, which gives all $S^{RL}_{2b}$ with $b=3,\cdots,n$ with $d=2$ in \eqref{SbootstrapLR}. Then, $S^{RL}_{23}$ determines $S^{RL}_{33}$, 
from which $S^{RL}_{3b}$ can be found for $b=4,\cdots,n$.  This procedure
continues until $S^{RL}_{nn}$.
The flow chart of determining all $S$-matrices looks like
\begin{eqnarray*}
S_{11}\to S_{12}\to S_{13}\to\cdots \to &S_{1n}&\\
\downarrow\hspace{3.1cm}& \\
S_{22}\to S_{23}\to \cdots \to &S_{2n}&\\
\downarrow\hspace{1.9cm}& \\
S_{33}\to\cdots \to &S_{3n}&\\
\cdots&\vdots\\
&S_{nn}&
\end{eqnarray*}
which reflects the Dynkin diagram structure in Fig.1 of the $A_n$ algebra.
Using $S^{RL}_{ab}=S^{RL}_{ba}$, all $S^{RL}_{ab}$ can be uniquely determined in this way from the initial ansatz for the $S^{RL}_{11}$.

Among $n^4$ relations in \eqref{SbootstrapLR} in a naive counting, about $n^2$ relations have been used in defining $S$-matrices.
All other relations should be considered as constraint equations for $S$-matrices.
In addition, the crossing-unitarity relations \eqref{crossunit}  should also be satisfied.
These constraints can determine the validity of $S^{RL}_{11}$.

Since all $S^{RL}_{ab}$ are determined uniquely by the initial ansatz $S^{RL}_{11}$, we can
classify the correct solutions of both the bootstrap equations and the crossing-unitarity relations by the $S^{RL}_{11}$. Although it would be tempting to use $T$ and $F_{\gamma}$ in \eqref{FTkernelRR}, it is not a good ansatz. 
To see this, let us consider $A_{2}$ case, whose bootstrap equations are particularly simple
\beq
S^{RL}_{ab}(\theta)S^{RL}_{ab}(\theta+\frac{i\pi}{3})S^{RL}_{ab}(\theta-\frac{i\pi}{3})=1,\qquad
a,b=1,2.
\label{a2bootstrap}
\eeq
It is easy to check that $T$ and $F_{\gamma}$ do not satisfy \eqref{a2bootstrap}. 
This implies that UV theory with the central charge $c_{\rm UV}=8/5$ in the list of
\cite{AhnLeC} is excluded by the bootstrap equations.

We have found that solutions $S^{RL}_{ab}$ which satisfy all constraints are
generated by 
\beq
S^{[m]}_{11}=\omega^{m_2}(m),\qquad m=1,2,\cdots,n,
\label{sii}
\eeq
where $(m)$ is defined in Eq.\eqref{defbasicblock} and $m_2=m\ {\rm mod}\ 2$. The phase $\omega=\exp(i\pi/(n+1))$ is introduced to satisfy the crossing-unitarity relation.
We claim that these and their products are the {\it only solutions}.
This observation is important for classifying UV-complete theories corresponding to $[A_n]_1$ IR CFT.

Furthermore, $S^{RL}$-matrices determined by the $S^{[m]}_{11}$ can be expressed as follows:
\beq
S_{ab}^{[m]}=\omega^{m_{2} ab} \prod_{k=1}^{{\rm min}(a,m)}
\left[\prod_{\ell=k}^{a+m-k}(a+b+m-2\ell)\right].
\label{masterSab}
\eeq
These $S$-matrices satisfy $S_{ab}^{[m]}=S_{ba}^{[m]}$ after simplifications using properties \eqref{defbasicblock}.

We have listed a few cases in detail.
\begin{enumerate}
\item $S^{[1]}_{11}(\theta)=\omega (1)$ : The $S$-matrices in \eqref{masterSab} are simplified to
\beq
S^{[1]}_{ab}=\omega^{ab} (a+b-1)(a+b-3)\cdots(|a-b|+1).
\label{smin}
\eeq
We will show that this set of $S$-matrices generates the ``minimal'' RG flow between $[A_n]_2$ and $[A_n]_1$ from a conjectured TBA \cite{Ravanini}.
To our knowledge, this is the first presentation of explicit $S$-matrices for these ``minimal'' RG flows.\footnote{This $S$-matrix has appeared in homogeneous sine-Gordon models \cite{Olalla} in a different context.}
We will derive a TBA system from these $S$-matrices in Section 3.

\item $S^{[n]}_{11}(\theta)=(n)=(\overline{1})$ : This generates a set of $S$-matrices which are related to the $S^{[1]}_{ab}$ by
\beq
S^{[n]}_{ab}=S^{[1]}_{a\overline{b}}.
\label{snewn}
\eeq
Therefore this also generates the RG flow from
$[A_n]_1$ to $[A_n]_2$.
This equivalence also holds for other values of $m$, that is,
$S^{[m]}_{ab}=S^{[\overline{m}]}_{a\overline{b}}$ for $m=1,\cdots,n$.

\item $S^{[2]}_{11}(\theta)=(2)$ : One can easily notice that Eq.\eqref{masterSab} becomes 
\beq
S^{[2]}_{ab}(\theta)=(a+b)(a+b-2)^2\cdots (|a-b|+2)^2 (|a-b|),
\label{masslessAnS}
\eeq
which is nothing but the algebraic inverse of the massive $S$-matrices
in \eqref{massiveAnS}. This case has been called ``diagonal'' and the RG flow connect $[A_n]_1$ to 
a parafermionic coset $su(n+1)_2/U(1)^n$ \cite{AhnLeC}. 

\item  $S^{[3]}_{11}(\theta)=\omega (3)$ :  Cases for $3\le m\le n-2$ with $n\ge 5$ are new sets of $S$-matrices which satisfy Eqs.\eqref{Sbootstrap} and \eqref{crossNunit}.
The simplest one among these is $m=3$, as follows: 
\bea
S^{[3]}_{1b}&=&S^{[3]}_{b1}=\omega^{b}(b+2)(b)(b-2),\\
S^{[3]}_{ab}&=&\omega^{ab}(a+b+1)(a+b-1)^2(a+b-3)^3(a+b-5)^3\cdots (|a-b|+3)^3\nonumber\\
&\times&(|a-b|+1)^2(|a-b|-1),\quad a,b\ge 2.
\label{snew3}
\eea
Although this set of $S$-matrices satisfies all conditions and is well-defined as IR scattering theories, we will show that it does not lead to any UV-complete theories.
\end{enumerate}

In summary, $S^{RL}$-matrices which satisfy the bootstrap equations and the crossing-unitarity relations are limited to
$S^{[m]}_{ab}$ in \eqref{masterSab} and their products such as
$S^{[m_1]}_{ab}S^{[m_2]}_{ab}\cdots$.
\section{Thermodynamic Bethe ansatz and RG flows}\label{sec3}
The TBA equations can be constructed from all these $S$-matrices, $S_{ab}^{RL}, S_{ab}^{LR}$,$S_{ab}^{LL}, and S_{ab}^{RR}$.
Since these are diagonal $S$-matrices, 
it is straightforward to derive the TBA equations
\bea
\epsilon^{R}_a(\theta)&=&\frac{\mu_a\mathbf{R}}{2}e^{\theta}-\sum_{b=1}^n\varphi_{ab}\star \mathbb{L}^R_b(\theta)-\sum_{b=1}^n\psi^{RL}_{ab}\star \mathbb{L}^L_b(\theta),
\label{rawtba}\\
\epsilon^{L}_a(\theta)&=&\frac{\mu_a\mathbf{R}}{2}e^{-\theta}-\sum_{b=1}^n\varphi_{ab}\star \mathbb{L}^L_b(\theta)-\sum_{b=1}^n\psi^{LR}_{ab}\star \mathbb{L}^R_b(\theta),\nonumber
\eea
with
\beq
\mathbb{L}^L_b=\log[1+e^{-\epsilon^L_b(\theta)}],\quad
\mathbb{L}^R_b=\log[1+e^{-\epsilon^R_b(\theta)}],
\eeq
and $\star$ represents the standard convolution.
The kernels of the TBA systems are given by
\bea
\varphi_{ab}(\theta)&=&-i\frac{d}{d\theta}\log S^{RR}_{ab}(\theta)=-i\frac{d}{d\theta}\log S^{LL}_{ab}(\theta),\\
\psi^{RL}_{ab}(\theta)&=&-i\frac{d}{d\theta}\log S^{RL}_{ab}(\theta),\quad
\psi^{LR}_{ab}(\theta)=-i\frac{d}{d\theta}\log S^{LR}_{ab}(\theta)=
\psi^{RL}_{ab}(-\theta),
\label{defkern}
\eea
where we have used the unitarity $S^{RL}(\theta)S^{LR}(-\theta)=1$ in the second line.
From this, it is obvious that $\epsilon_a^R(\theta)=\epsilon_a^L(-\theta)$.
We consider each case of $S^{RL}=S^{[m]}$ in detail as follows.

\subsection{Minimal flows ($m=1$ or $m=n$)}
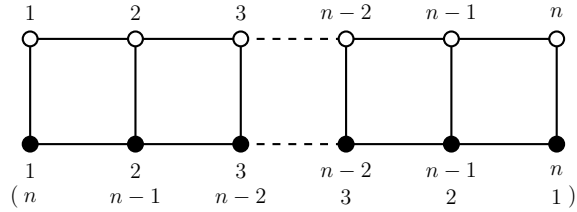
\begin{figure}[h]
\centering
\begin{tikzpicture}[scale=.7, transform shape]
\draw[thick] (1,0) -- (3,0);
\draw[thick] (3,0) -- (5,0);
\draw[thick,dashed] (5,0) -- (7,0);
\draw[thick] (7,0) -- (9,0);
\draw[thick] (9,0) -- (11,0);
\draw[thick] (1,2) -- (3,2);
\draw[thick] (3,2) -- (5,2);
\draw[thick,dashed] (5,2) -- (7,2);
\draw[thick] (7,2) -- (9,2);
\draw[thick] (9,2) -- (11,2);
\draw[thick] (1,0) -- (1,2);
\draw[thick] (3,0) -- (3,2);
\draw[thick] (5,0) -- (5,2);
\draw[thick] (7,0) -- (7,2);
\draw[thick] (9,0) -- (9,2);
\draw[thick] (11,0) -- (11,2);
\filldraw[color=black, fill=black, thick] (1,0) circle (4pt);
\filldraw[color=black, fill=black, thick] (3,0) circle (4pt);
\filldraw[color=black, fill=black, thick] (5,0) circle (4pt);
\filldraw[color=black, fill=black, thick] (7,0) circle (4pt);
\filldraw[color=black, fill=black, thick] (9,0) circle (4pt);
\filldraw[color=black, fill=black, thick] (11,0) circle (4pt);
\filldraw[color=black, fill=white, thick] (1,2) circle (4pt);
\filldraw[color=black, fill=white, thick] (3,2) circle (4pt);
\filldraw[color=black, fill=white, thick] (5,2) circle (4pt);
\filldraw[color=black, fill=white, thick] (7,2) circle (4pt);
\filldraw[color=black, fill=white, thick] (9,2) circle (4pt);
\filldraw[color=black, fill=white, thick] (11,2) circle (4pt);
\node [label] at (1,-.5) {$1$};
\node [label] at (3,-.5) {$2$};
\node [label] at (5,-.5) {$3$};
\node [label] at (11,-.5) {$n$};
\node [label] at (9,-.5) {$n-1$};
\node [label] at (7,-.5) {$n-2$};
\node [label] at (0.7,-1) {$($};
\node [label] at (1,-1) {$n$};
\node [label] at (3,-1) {$n-1$};
\node [label] at (5,-1) {$n-2$};
\node [label] at (11,-1) {$1$};
\node [label] at (9,-1) {$2$};
\node [label] at (7,-1) {$3$};
\node [label] at (11.3,-1) {$)$};
\node [label] at (1,2.5) {$1$};
\node [label] at (3,2.5) {$2$};
\node [label] at (5,2.5) {$3$};
\node [label] at (11,2.5) {$n$};
\node [label] at (9,2.5) {$n-1$};
\node [label] at (7,2.5) {$n-2$};
\end{tikzpicture}
\caption{Universal TBA for the minimal RG flows of $A_n$ algebra. White (Black) nodes are for $R(L)$ pseudoenergies for $m=1$. Indices in the parenthesis are for $m=n$.}
\label{fig2}
\end{figure}
For analytic manipulations of the TBA, it is convenient to use Fourier transformed kernels.
The basic building block of the $S$-matrices $(x)$ in \eqref{defbasicblock} contributes
\beq
\phi_{k}(\theta)=-i\frac{d}{d\theta}\log (k)\quad \to \quad
\widetilde{\phi}_k={\rm sgn}(k)\frac{\sinh w\left(1-\frac{|k|}{n+1}\right)}{\sinh w},
\label{basicFT}
\eeq
where ${\rm sgn}(k)=1,0,-1$ for $k>0,k=0,k<0$, respectively.
Using this, 
the Fourier transformed kernels for the $RL$ and $LR$ sectors with $m=1,n$
are obtained as\footnote{We denote the Fourier transform of a function $f$ with $\widetilde{f}$.}
\bea
\widetilde{\psi}^{[1]}_{ab}(w)&=&\frac{\sinh\left(\frac{aw}{h}\right)\sinh\left(\frac{\overline{b}w}{h}\right)}{\sinh w\sinh\left(\frac{w}{h}\right)},\\
\widetilde{\psi}^{[n]}_{ab}(w)&=&\frac{\sinh\left(\frac{a'w}{h}\right)\sinh\left(\frac{b'w}{h}\right)}{\sinh w\sinh\left(\frac{w}{h}\right)},\qquad (a',b')=
\begin{cases}
(a,b),&a+b\le n+1\\
(\overline{a},\overline{b}),&a+b> n+1.
\end{cases}
\eea
The kernels in the $RR$ and $LL$ sectors are given by
\beq
\widetilde{\varphi}_{ab}(w)=-\widetilde{\psi}^{[2]}_{ab}(w)
=\delta_{a,b}-\frac{\sinh\left(\frac{aw}{h}\right)\sinh\left(\frac{\overline{b}w}{h}\right)\sinh\left(\frac{2w}{h}\right)}{\sinh w\sinh^2\left(\frac{w}{h}\right)}.
\label{FTkernelRR}
\eeq
In these kernels, $b\ge a$ is assumed. If not, $a\leftrightarrow b$ can be 
applied.

A well-known relation for
$\widetilde{\varphi}_{ab}(w)$ is \cite{ZamADE}
\beq
{\left[\mathbf{1}-\widetilde{\varphi}(w)\right]^{-1}}_{ab}=
\delta_{ab}-\widetilde{\phi}_h(w)\mathbb{I}_{ab},\qquad
\widetilde{\phi}_h(w)=\frac{1}{2\cosh\frac{w}{h}}
\label{alyosha}
\eeq
where $\mathbf{1}$ is the identity matrix, $\mathbb{I}$ is the incidence matrix of $A_n$ in Fig.\ref{fig1}.
Other useful relations are 
\beq
{\left[\mathbf{1}-\widetilde{\varphi}(w)\right]^{-1}}\cdot 
\widetilde{\psi}^{[1]}(w)=\widetilde{\phi}_h(w)\mathbf{1},\qquad
{\left[\mathbf{1}-\widetilde{\varphi}(w)\right]^{-1}}\cdot 
\widetilde{\psi}^{[n]}(w)=\widetilde{\phi}_h(w)\overline{\mathbf{1}}
\eeq
with $\overline{\mathbf{1}}_{ab}=\delta_{a\overline{b}}$.
Using these relations, we can convert the raw TBA equations \eqref{rawtba} to the universal TBAs,
\bea
\epsilon^{R}_a(\theta)&=&\frac{\mu_a\mathbf{R}}{2}e^{\theta}-\phi_{h}\star\left[\sum_{b=1}^n\mathbb{I}_{ab}\, \overline{\mathbb{L}}^R_b(\theta)-\mathbb{L}_{a'}^{L}
\right](\theta),\\
\epsilon^{L}_a(\theta)&=&\frac{\mu_a\mathbf{R}}{2}e^{-\theta}-\phi_{h}\star\left[\sum_{b=1}^n\mathbb{I}_{ab}\, \overline{\mathbb{L}}^L_b(\theta)-\mathbb{L}_{a'}^{R}
\right](\theta),
\eea
where $a'=a$ for $m=1$, $a'=\overline{a}$ for $m=n$, and
\beq
\overline{\mathbb{L}}^{L/R}_b =\log(1+e^{\epsilon^{L/R}_b}).
\eeq

These can also be expressed in terms of the $Y$-system as follows:
\beq
Y^{R}_a(\theta+\frac{i\pi}{h})Y^{R}_a(\theta-\frac{i\pi}{h})
=\frac{\displaystyle{\prod_{b=1}^n}(1+Y^{R}_b(\theta))^{\mathbb{I}_{ab}}}
{(1+1/Y^{L}_{a'}(\theta))},\quad
Y^{L}_a(\theta+\frac{i\pi}{h})Y^{L}_a(\theta-\frac{i\pi}{h})
=\frac{\displaystyle{\prod_{b=1}^n}(1+Y^{L}_b(\theta))^{\mathbb{I}_{ab}}}
{(1+1/Y^{R}_{a'}(\theta))}.
\eeq

One application of the $Y$-system is to find the periodicity 
$Y_a^{R/L}(\theta+2\pi i T)=Y_a^{R/L}(\theta)$ where $T$ is related to
the dimension of the relevant operator $\Delta_{\rm rel}$ by
$T=2/(2-\Delta_{\rm rel})$.
Above $Y$-system produces the periodicity to be $(h+3)/h$, which means
$\Delta_{\rm rel}=6/(h+3)$.
Comparing with \eqref{dimrel}, we can confirm that the UV theory is indeed the theory in Eq.\eqref{deformedcft} with $G=A_n$ and $k=2$.

\subsection{Diagnonal flows for $m=2$ or $m=n-1$}
From Eq.\eqref{FTkernelRR}, it is straightforward to write the $Y$-system 
\beq
Y^{R}_a(\theta+\frac{i\pi}{h})Y^{R}_a(\theta-\frac{i\pi}{h})
=\displaystyle{\prod_{b=1}^n}\left[\frac{1+Y^{R}_b(\theta))}{1+1/Y^{L}_{b}(\theta)}\right]^{\mathbb{I}_{ab}},\quad
Y^{L}_a(\theta+\frac{i\pi}{2})Y^{L}_a(\theta-\frac{i\pi}{2})
=\displaystyle{\prod_{b=1}^n}\left[\frac{1+Y^{L}_b(\theta))}{1+1/Y^{R}_{b}(\theta)}\right]^{\mathbb{I}_{ab}}.
\eeq
\begin{figure}[h]
\centering
\begin{tikzpicture}[scale=.7, transform shape]
\draw[thick] (1,0) -- (3,0);
\draw[thick] (3,0) -- (5,0);
\draw[thick] (7,0) -- (9,0);
\draw[thick] (9,0) -- (11,0);
\draw[thick] (1,2) -- (3,2);
\draw[thick] (3,2) -- (5,2);
\draw[thick] (7,2) -- (9,2);
\draw[thick] (9,2) -- (11,2);

\draw[thick,dashed] (1,0) -- (3,2);
\draw[thick,dashed] (3,2) -- (5,0);
\draw[thick,dashed] (5,0) -- (7,2);
\draw[thick,dashed] (7,2) -- (9,0);
\draw[thick,dashed] (9,0) -- (11,2);
\draw[thick,dashed] (1,2) -- (3,0);
\draw[thick,dashed] (3,0) -- (5,2);
\draw[thick,dashed] (5,0) -- (7,0);
\draw[thick,dashed] (5,2) -- (7,2);
\draw[thick,dashed] (5,2) -- (7,0);
\draw[thick,dashed] (7,0) -- (9,2);
\draw[thick,dashed] (9,2) -- (11,0);
\filldraw[color=black, fill=black, thick] (1,0) circle (4pt);
\filldraw[color=black, fill=black, thick] (3,0) circle (4pt);
\filldraw[color=black, fill=black, thick] (5,0) circle (4pt);
\filldraw[color=black, fill=black, thick] (7,0) circle (4pt);
\filldraw[color=black, fill=black, thick] (9,0) circle (4pt);
\filldraw[color=black, fill=black, thick] (11,0) circle (4pt);
\filldraw[color=black, fill=white, thick] (1,2) circle (4pt);
\filldraw[color=black, fill=white, thick] (3,2) circle (4pt);
\filldraw[color=black, fill=white, thick] (5,2) circle (4pt);
\filldraw[color=black, fill=white, thick] (7,2) circle (4pt);
\filldraw[color=black, fill=white, thick] (9,2) circle (4pt);
\filldraw[color=black, fill=white, thick] (11,2) circle (4pt);
\node [label] at (1,-.5) {$1$};
\node [label] at (3,-.5) {$2$};
\node [label] at (5,-.5) {$3$};
\node [label] at (11,-.5) {$n$};
\node [label] at (9,-.5) {$n-1$};
\node [label] at (7,-.5) {$n-2$};
\node [label] at (0.7,-1) {$($};
\node [label] at (1,-1) {$n$};
\node [label] at (3,-1) {$n-1$};
\node [label] at (5,-1) {$n-2$};
\node [label] at (11,-1) {$1$};
\node [label] at (9,-1) {$2$};
\node [label] at (7,-1) {$3$};
\node [label] at (11.3,-1) {$)$};
\node [label] at (1,2.5) {$1$};
\node [label] at (3,2.5) {$2$};
\node [label] at (5,2.5) {$3$};
\node [label] at (11,2.5) {$n$};
\node [label] at (9,2.5) {$n-1$};
\node [label] at (7,2.5) {$n-2$};
\end{tikzpicture}
\caption{Universal TBA for the diagonal RG flows of $A_n$ algebra for $m=2$. Indices in the parenthesis are for $m=n-1$.}
\label{fig3}
\end{figure}
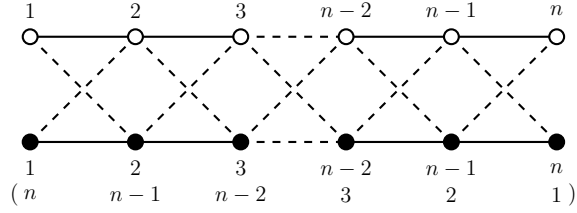

This TBA system, depicted by Fig.\ref{fig3},
should generate an RG flow from 
$[A_n]_1$ to 
a parafermionic coset $su(n+1)_2/U(1)^n$ 
as discovered in \cite{AhnLeC}. 
It turns out that the periodicity of the $Y$ functions is infinite 
which means the UV theory is deformed by some marginal operator
with $\Delta=2$.
For more details, we need to analyse the
$Y$-system in the UV domain, which is not pursued here.

\subsection{No UV-complete theories for $3\le m\le n-2$}
We analysed these cases by solving the plateau equations numerically.
For example, we consider $[A_5]_1$, which is the first case of $m=3$.
The $\widetilde{\varphi}_{ab}(0)$ and
$\widetilde{\psi}^{[3]}_{ab}(0)$ are given by
\beq
\widetilde{\varphi}_{ab}(0)=-\frac{1}{3}\left(
\begin{array}{ccccc}
2&4&3&2&1\\ 4&5&6&4&2\\3&6&6&6&3\\
2&4&6&5&4\\1&2&3&4&2
\end{array}
\right),\qquad
\widetilde{\psi}^{[3]}_{ab}(0)=\frac{1}{2}\left(
\begin{array}{ccccc}
1&2&3&2&1\\ 2&4&4&4&2\\3&4&5&4&3\\
2&4&4&4&2\\1&2&3&2&1
\end{array}
\right).\qquad
\eeq
Because the exponents of the plateau equations, given by the sum of these two matrices, are mostly positive, the equations do not have finite real solutions.
We have checked sufficiently many cases in this type by solving the plateau equations numerically and found that no finite rational central charges are produced. 
Although this kind of case studies cannot be a mathematically rigorous proof, we claim that 
no RG flows that lead to UV-complete theories can be obtained.
This implies that the Hagedorn singularity arises before reaching the UV limit.

\subsection{Saturated flows for $m=1$ \& $m=n$}
It is possible to find additional solutions to the bootstrap equations
by combining some of the basic solutions denoted by index $m$.
We have confirmed that the only a product of the $m=1$ and $m=n$
$S^{RL}$-matrices,
\beq
S^{RL}_{ab}(\theta)=S^{[1]}_{ab}(\theta)S^{[n]}_{ab}(\theta),
\label{Ssaturated}
\eeq
can reach the UV limit and reproduce the saturated flows discovered in \cite{AhnLeC}.

The $Y$-system can be obtained from the universal TBA equations for this case 
\beq
Y^{R}_a(\theta+\frac{i\pi}{h})Y^{R}_a(\theta-\frac{i\pi}{h})
=\frac{\displaystyle{\prod_{b=1}^n}(1+Y^{R}_b(\theta))^
{\mathbb{I}_{ab}}}
{(1+1/Y^{L}_{a}(\theta))(1+1/Y^{L}_{\overline{a}}(\theta))},
\eeq
and similarly for $R\leftrightarrow L$.

One can check that this generates the saturated RG flows from $[A_n]_1$ to 
$su(n+1)_2$ WZW theory.
Similar to the diagonal cases, the UV theory also has only marginal deformation. 
More information on this theory can be identified by comparing a detailed analysis of the above 
$Y$-system with the result based on the marginal operator in the UV domain.
We hope to report on this in the near future.

\section{Summary}\label{sec5}
We have applied the massless $S$-matrix bootstrap equations for the $[A_n]_1$ theory and found systematic solutions for them.
Each solution starts with $S_{11}^{RL}=(m)$ and all $S$-matrices 
between the left- and right-moving particles are defined by some bootstrap equations. 
It has been confirmed that only those with $m=1,\cdots,n$ satisfy the rest of the equations and crossing-unitarity relations.
In addition, the products of these basic solutions satisfy all equations. 

Only some of these solutions generate RG flows that can reach UV-complete theories from $[A_n]_1$ coset CFTs.
The $S^{RL}$-matrices that  produce the RG flows leading to UV-complete theories are
only the minimal, diagonal, and saturated solutions.
\beq
{\rm minimal:}\ S_{ab}^{[1]}\ \& S_{ab}^{[n]},\quad
{\rm diagonal:}\ S_{ab}^{[2]}\ \& S_{ab}^{[n-1]},\quad
{\rm saturated:}\ S_{ab}^{[1]}\cdot S_{ab}^{[n]}
\label{summary}
\eeq
Most other $S$-matrix theories fail at some scale due to the occurrence of the Hagedorn singularity. 
Among the many potential UV theories predicted in a previous analysis \cite{AhnLeC} based on the plateau equations, but without imposing the bootstrap conditions, are excluded.

We have also derived the exact TBA equations and corresponding $Y$-systems to analyse these UV-complete theories in more detail.
We have proved explicitly the RG flows between
$[A_n]_2$ and $[A_n]_1$ along definite relevant and irrelevant directions, respectively, using the periodicities of the $Y$-systems.
Although these flows have been known for a long time, we believe that this is the first rigorous derivation based on exact $S$-matrices.

All the above conclusions are limited to the $G=A_n$ coset theories. 
The method used in this paper can be applied to other groups, such as $G=D,\ E$ and their cosets.
Particularly interesting is $[E_8]_1$ since it is another representation of the Ising model with the central charge $1/2$.
It will be interesting to find all $S^{RL}$-matricx solutions to the bootstrap equations and check if new types of RG flows other than those in \eqref{summary} can be allowed. 
We hope to report on this in the near future.

Recently, RG flows between nonunitary CFTs have been studied in the context of noninvertible symmetries \cite{NakTan}. It would be interesting to investigate these from our massless $S$-matrix approach.

\section*{Acknowledgements}
We thank Z. Bajnok and A. LeClair for their valuable discussions, useful comments, careful reading and encouragement. 
This work was supported by the National Research Foundation of Korea (NRF) grant
(NRF-2016R1D1A1B02007258).


\begin{thebibliography}{99}
\bibitem{sasha} A. B. Zamolodchikov, {\em Integrals of motions and S-matrix of the (scaled) T=Tc Ising model with magnetic field } , Int. J. Mod. Phys. {\bf A4} (1989) 4235.

\bibitem{alyosha_tba} Al. B. Zamolodchikov,  {\em Thermodynamic Bethe ansatz in Relativistic Models: Scaling 3-state Potts and Lee-Yang Models}, Nucl. Phys. {\bf B342} (1990) 695.

\bibitem{zamolRG}  A. B. Zamolodchikov, 
{\em Irreversibility of the flux of the renormalization group in a 2D field theory},
Pis'ma Eksp. Teor. Fiz. {\bf 43} (1986) 565.

\bibitem{alyosha_tba_flow}Al. B. Zamolodchikov, {\em From Tricritical Ising to Critical Ising By Thermodynamic Bethe ansatz}, Nucl. Phys. {\bf B358} (1991) 524.

\bibitem{alyosha_tba_cosetflow}Al.B. Zamolodchikov, {\em TBA Equations for Integrable Perturbed $SU(2)_k\times  SU(2)_l/SU(2)_{k+l}$ Coset Models}, Nucl. Phys. {\bf B366} (1991) 122.

\bibitem{Ravanini} F. Ravanini, 
{\it Theormodynamic Bethe Ansatz for $G_k \otimes G_\ell / G_{k + \ell}$ coset models perturbed by their $\phi_{1,1,{\rm Adj}} $ operator}, 
Phys. Lett. {\bf B282} (1992) 73, arXiv:hep-th/9202020.

\bibitem{DDT}P. Dorey, C. Dunning and R. Tateo, {\em New families of flows between two-dimensional conformal field theories}, Nucl. Phys. {\bf B578} (2000) 699, arXiv:0001185 [hep-th]. 

\bibitem{zamzam_flow}A. B. Zamolodchikov and Al.B. Zamolodchikov, {\em Massless factorized scattering and sigma models with topological terms}, Nucl. Phys. {\bf B379} (1992) 602.

\bibitem{FSZ} P. Fendley, H. Saleur, and Al. B. Zamolodchikov, {\em Massless Flows II: the exact S-matrix approach}, Int. J. Mod. Phys. {\bf A8} (1993) 5751, arXiv:hep-th/930405.

\bibitem{AKRZ}C. Ahn, C. Kim, C. Rim, and Al.B. Zamolodchikov, {\em RG flows from super-Liouville theory to critical Ising model}, Phys. Lett. {\bf B541} (2002) 194.

\bibitem{AhnLeC} C. Ahn and A. LeClair, {\em On the classification of UV completions of integrable $ T{\overline T}$  deformations of CFT}, JHEP {\bf 2022} (2022) 179, arXiv:2205.10905 [hep-th].

\bibitem{SmiZam} F. A. Smirnov and A. B. Zamolodchikov,  {\em On the space of integrable quantum field theories}, Nucl. Phys. {\bf  B915}  (2017) 363, arXiv:1608.05499 [hep-th].

\bibitem{AhnBaj} C. Ahn and Z. Bajnok, {\em New integrable RG flows with parafermions}, JHEP {\bf 2411} (2024) 078, arXiv: 2407.06582 [hep-th].

\bibitem{BCDS} H. W. Braden, E. Corrigan, P. E. Dorey, and R. Sasaki, 
{\em Affine Toda field theory and exact S-matrices},  Nucl. Phys. {\bf B338} (1990) 689.

\bibitem{ABL} C. Ahn, D. Bernard and A. LeClair, {\em Fractional Supersymmetries in Perturbed Coset CFTs and Integrable Soliton Theory}, Nucl. Phys. {\bf B346} (1990) 409.

\bibitem{Olalla} O.A. Castro-Alvaredo, A. Fring, C. Korff and J.L. Miramontes, 
{\em Thermodynamic Bethe ansatz of the homogeneous sine-Gordon models},
Nucl. Phys. {\bf B575} (2000) 535, arXiv:hep-th/9912196. 

\bibitem{ZamADE} Al.B. Zamolodchikov, {\em On the thermodynamic Bethe ansatz 
equations for reflectionless ADE scattering theories}, Phys. Lett. {\bf B253} (1991) 391.

\bibitem{NakTan}  Y. Nakayama and T. Tanaka, {\em Infinitely many new renormalization group flows between Virasoro minimal models from non-invertible symmetries}, JHEP {\bf 2411} (2024) 137, arXiv:2407.21353 [hep-th].





\end{thebibliography}
\end{document}